\documentclass[10pt]{article}
\usepackage{amsmath,epsfig,multicol,bbm,times,graphicx,amssymb}

\begin{document}

\def\cL{\mathcal{L}}
\def\cA{\mathcal{A}}
\def\cM{\mathcal{M}}
\def\cN{\mathcal{N}}
\def\cP{\mathcal{P}}
\def\cR{\mathcal{R}}
\def\fM{\mathfrak{M}}
\def\fA{\mathfrak{A}}
\def\IR{\mathbb{R}}
\def\ZZ{\mathbb{Z}}
\def\todo{{\bf --- below not edited ---}}
\def\pnote#1{{\bf P: #1}}
\def\tnote#1{{\bf T: #1}}
\def\enote#1{{\bf E: #1}}
\def\phnote#1{{\bf PH: #1}}

\def\tr{\textrm{tr}}
\def\cL{\mathcal{L}}
\def\cA{\mathcal{A}}
\def\cM{\mathcal{M}}
\def\cN{\mathcal{N}}
\def\cP{\mathcal{P}}
\def\cR{\mathcal{R}}
\def\fM{\mathfrak{M}}
\def\fA{\mathfrak{A}}
\def\IR{\mathbb{R}}
\def\ZZ{\mathbb{Z}}
\def\enote#1{{\bf E: #1}}
\def\pnote#1{\textbf{P: #1}}
\def\todo{TO BE EDITED}
\def\stamp{--- {\bf \today} --- {\bf \jobname.tex}}

\def\A#1#2{\la#1#2\ra}
\def\B#1#2{[#1#2]}
\def\AB#1#2#3{\la#1|#2|#3]}
\def\BA#1#2#3{[#1|#2|#3\ra}
\def\AA#1#2#3{\la#1|#2|#3\ra}
\def\BB#1#2#3{[#1|#2|#3]}

\def\fA{\mathfrak{A}}
\def\fB{\mathfrak{B}}
\def\fI{\mathfrak{I}}
\def\fS{\mathfrak{S}}
\def\fG{\mathfrak{G}}
\def\cG{\mathcal{G}}
\def\cB{\mathcal{B}}

\def\tr{\textrm{tr}}
\def\cSS#1#2{{\mathcal{S}}_{{#1},{#2}}}
\def\ss#1#2{s_{#1,#2}}

\def\kk{\frac{\kappa_{(4)}}{2}}

\def\todo{{\bf --- below not edited ---}}
\def\more#1{\footnote{\bf more: #1}}
\def\cN{\mathcal{N}}
\def\cP{\mathcal{P}}
\def\cR{\mathcal{R}}
\def\BE{\begin{equation}}
\def\EE{\end{equation}}
\def\spa#1.#2{\left\langle#1\,#2\right\rangle}
\def\spb#1.#2{\left[#1\,#2\right]}
\def\spba#1.#2.#3{\left[#1|#2|#3\right\rangle}
\def\spab#1.#2.#3{\left\langle#1|#2|#3\right]}
\def\spaa#1.#2.#3{\left\langle#1|#2|#3\right\rangle}
\def\spbb#1.#2.#3{\left[#1|#2|#3\right]}
\def\lor#1.#2{\left(#1\,#2\right)}
\def\sign{{\mathop{\rm sign}\nolimits}}

\def\Year{\expandafter\eatPrefix\the\year}
\newcount\hours \newcount\minutes
\def\monthname{\ifcase\month\or
January\or February\or March\or April\or May\or June\or July\or
August\or September\or October\or November\or December\fi}
\def\shortmonthname{\ifcase\month\orx
Jan\or Feb\or Mar\or Apr\or May\or Jun\or Jul\or Aug\or Sep\or
Oct\or Nov\or Dec\fi}
\def\thedate{\monthname\space\number\day, \number\year}
\def\TimeStamp{\hours\the\time\divide\hours by60%
\minutes -\the\time\divide\minutes by60\multiply\minutes by60%
\advance\minutes by\the\time%
${\rm \shortmonthname}\cdot   \if\day<10{}0\fi\the\day\cdot
\the\year \qquad\the\hours:\if\minutes<10{}0\fi\the\minutes$}

\newcommand{\figmac}[4]{\begin{figure}
\centerline{\parbox[t]{#1in}{\epsfbox{#2.eps}}}
\caption[#4]{\label{fig:#3} #4}\end{figure}}
\newskip\humongous \humongous=0pt plus 1000pt minus 100pt
\def\caja{\mathsurround=0pt}
\def\eqalign#1{\,\vcenter{\openup1\jot \caja
       \ialign{\strut \hfil$\displaystyle{##}$&$
        \displaystyle{{}##}$\hfil\crcr#1\crcr}}\,}
\newif\ifdtup
\def\panorama{\global\dtuptrue \openup1\jot \caja
        \everycr{\noalign{\ifdtup \global\dtupfalse
        \vskip-\lineskiplimit \vskip\normallineskiplimit
        \else \penalty\interdisplaylinepenalty \fi}}}
\def\eqalignno#1{\panorama \tabskip=\humongous
        \halign to\displaywidth{\hfil$\displaystyle{##}$
        \tabskip=0pt&$\displaystyle{{}##}$\hfil
        \tabskip=\humongous&\llap{$##$}\tabskip=0pt
        \crcr#1\crcr}}

%
\newcounter{eqnumber}[section]
\renewcommand{\theeqnumber}{\thesection.\arabic{eqnumber}}
\def\equn{\refstepcounter{eqnumber}
\eqno({\rm \theeqnumber}) }
\def\equnno{\refstepcounter{eqnumber}
({\rm \theeqnumber})}

\def\npb#1#2#3{{\rm Nucl. Phys. B}{\bf \ #1}, #3 (#2)}
\def\plb#1#2#3{{\rm Phys. Lett. B}{\bf \ #1}, #3 (#2)}
\def\prd#1#2#3{{\rm Phys. Rev. D} {\bf  #1}, #3 (#2)}
\def\cqg#1#2#3{{\rm Class. and Quant.\ Grav.} {\bf  #1}, #3 (#2)}
\def\prl#1#2#3{{\rm Phys. Rev. Lett. } {\bf  #1}, #3 (#2)}
\def\JHEP#1#2#3{{JHEP  } {\bf  #1}, #3 (#2)}
\def\JHEPproc#1#2#3{{\it J.\ High\ Ener.\ Phys.\  } {\bf  #1}:#3 (#2)}
\def\ijmp#1#2#3{{\rm Int.\ J.\ Mod.\ Phys.\ } {\bf A #1}, #3 (#2)}
\def\jmp#1#2#3{{\rm  J. Math.\ Phys.\ } {\bf  #1}, #3 (#2)}
\def\hepth#1{[hep-th/#1]}
\def\hepph#1{[hep-ph/#1]}


\def\qb{{\bar q}}
\def\eqn#1{eq.~(\ref{#1})}
\def\Eqn#1{Eq.~(\ref{#1})}
\def\eqns#1#2{eqs.~(\ref{#1}) and~(\ref{#2})}
\def\Eqns#1#2{Eqs.~(\ref{#1}) and~(\ref{#2})}
\def\fig#1{fig.~#1}
\def\figs#1#2{figs.~{\ref{#1}} and {\ref{#2}}}
\def\Figs#1#2{Figs.~{\ref{#1}} and {\ref{#2}}}
\def\sec#1{section~{\ref{#1}}}
\def\Sec#1{Section~{\ref{#1}}}
\def\secs#1#2{sections~{\ref{#1}} and {\ref{#2}}}
\def\app#1{appendix~\ref{#1}}
\def\App#1{Appendix~\ref{#1}}
\def\apps#1#2{appendices~{\ref{#1}} and {\ref{#2}}}
\def\Apps#1#2{Appendices~{\ref{#1}} and {\ref{#2}}}
\def\tab#1{table~\ref{#1}}
\def\Tab#1{Table~\ref{#1}}
\def\tr{\mathop{\rm tr}\nolimits}
\def\dilog{\mathop{\rm dilog}\nolimits}
\def\trplus{\mathop{\rm tr}\nolimits_+}
\def\trminus{\mathop{\rm tr}\nolimits_-}

\def\Is#1#2{F
^{{#1}}_{4:#2}}
\def\Ione{\Is{\rm 1m}}
\def\Ieasy{\Is{{\rm 2m}\,e}}
\def\Ihard{\Is{{\rm 2m}\,h}}

\def\Fn{n}
\def\Fs#1#2{F^{{#1}}_{\Fn:#2}}
\def\Fone{\Fs{\rm 1m}}
\def\Feasy{\Fs{{\rm 2m}\,e}}\def\Feasytwo{\Feasy}
\def\Fhard{\Fs{{\rm 2m}\,h}}\def\Fhardtwo{\Fhard}
\def\Fthree{\Fs{\rm 3m}}
\def\Ffour{\Fs{\rm 4m}}

\newbox\charbox
\newbox\slabox

\def\spa#1.#2{\left\langle#1\,#2\right\rangle}
\def\spb#1.#2{\left[#1\,#2\right]}
\def\lor#1.#2{\left(#1\,#2\right)}

\catcode`@=11  

\def\Slash#1{\hskip 0.05 cm \slash\hskip -0.22 cm #1}
\def\as#1{a_{\sigma(#1)}}
\def\sig#1{\sigma(#1)}
\def\si{\sigma}
\def\Tr{\, {\rm Tr}}
\def\NP{{\rm NP}}
\def\LC{{\rm LC}}
\def\SC{{\rm SC}}
\def\eps{\epsilon}
\def\alp{\alpha}
\def\Ord{{\cal O}}
\def\neqfour{N=4}
\def\scut{{s\rm \hbox{-}cut}}
\def\I{{\cal I}}
\def\fourperm#1#2#3#4{(#1\,#2\,#3\,#4)}%
\def\susy{{\scriptscriptstyle \rm SUSY}}
\def\pol{\eps}
\def\zb{\bar z}
\def\x#1#2{x_{#1 #2}}
\def\gT{{\tilde g}}
\def\hT{{\tilde h}}
\def\vT{{\tilde v}}
\def\e{\epsilon}
\def\lr{\leftrightarrow}

\def\half{{1\over 2}}
\def\threehalf{{3\over 2}}
\def\la{\langle}
\def\ra{\rangle}
\def\oneloop{{1 \mbox{-} \rm loop}}
\def\twoloop{{2 \mbox{-} \rm loop}}
\def\Lloop{{L \mbox{-} \rm loop}}
\def\sumYM{\sum_{S_1, S_2 \in \{N=4\} }}
\def\sumGrav{\sum_{S_1, S_2 \in \{N=8\} }}
\def\Sp{{\hbox{\tiny$S^\prime$}}}
\def\vp{\vphantom{\Big|}}
\def\lsl{\not{\hbox{\kern-2.3pt $\ell$}}}
\def\ksl{\not{\hbox{\kern-2.3pt $k$}}}

\def\bvec#1{${\bf W}{}_{#1}$}
\def\H{\half}
\def\Z{0}
\def\Q{\quart}
\def\rg{r_{\Gamma}}

\def\spa#1.#2{\left\langle#1\,#2\right\rangle}
\def\spb#1.#2{\left[#1\,#2\right]}
\def\lor#1.#2{\left(#1\,#2\right)}

\def\sand#1.#2.#3{%
  \left\langle\smash{#1}{\vphantom1}\right|{#2}%
  \left|\smash{#3}{\vphantom1}\right\rangle}
\def\sandp#1.#2.#3{%
  \left\langle\smash{#1}{\vphantom1}^{-}\right|{#2}%
  \left|\smash{#3}{\vphantom1}^{+}\right\rangle}
\def\sandpp#1.#2.#3{%
  \left\langle\smash{#1}{\vphantom1}^{+}\right|{#2}%
  \left|\smash{#3}{\vphantom1}^{+}\right\rangle}
\def\sandmm#1.#2.#3{%
  \left\langle\smash{#1}{\vphantom1}^{-}\right|{#2}%
  \left|\smash{#3}{\vphantom1}^{-}\right\rangle}
\def\sandpm#1.#2.#3{%
  \left\langle\smash{#1}{\vphantom1}^{+}\right|{#2}%
  \left|\smash{#3}{\vphantom1}^{-}\right\rangle}
\def\sandmp#1.#2.#3{%
  \left\langle\smash{#1}{\vphantom1}^{-}\right|{#2}%
  \left|\smash{#3}{\vphantom1}^{+}\right\rangle}

\def\Aloop{A^{\rm 1-loop}}

\def\ACalloop{{\cal A}^{\rm 1-loop}}

\def\Mloop{M^{\rm 1-loop}}
\def\Mtree{M^{\rm tree}}
\def\Mcount{M^{\rm coun.}}
\def\Atree{A^{\rm tree}}
\def\Aloop{A^{\rm 1-loop}}
\def\A#1#2{\left\langle #1|#2\right\rangle}
\def\ke#1#2{ \eps_{#2} \cdot k_{#1} }
\def\mom#1#2{ k_{#1} \cdot k_{#2}}
\def\gbd{{\dot G}}
\def\gbdd{{\ddot G}}
\def\gbdbar{ { \overline{\dot G}} }
\def\gbdb{\gbdbar}
\def\gbddbar{{ \overline{\ddot G}}}
\def\dlips{dLIPS}

\def\Lz{\mathop{\hbox{\rm L}}\nolimits_0}
\def\Kz{\mathop{\hbox{\rm K}}\nolimits_0}
\def\Mz{\mathop{\hbox{\rm M}}\nolimits_0}
\def\tr{\mathop{\hbox{\rm tr}}\nolimits}

\def\L{\left(}\def\R{\right)}
\def\LP{\left.}\def\RP{\right.}
\def\LB{\left[}\def\RB{\right]}
\def\LA{\left\langle}\def\RA{\right\rangle}
\def\LV{\left|}\def\RV{\right|}

\def\polk#1#2{\eps_{#1}\cdot k_{#2}}

\def\ksumA#1{\biggl(\sum_i \eps_{#1} \cdot k_i \gbd_{#1,i}\biggr)}
\def\ksumB#1{\biggl(\sum_i k_{#1} \cdot k_i \gbd_{#1 , i}\biggr) }
\def\ksumC#1#2{\biggl(
\sum_i \Bigl( k_{#1} \cdot k_i \gbd_{#1 , i} +
 k_{#2} \cdot k_i \gbd_{#2 , i} \Bigr) \biggr) }
\def\sume#1{\left( \sum_{i} \eps_{#1} \cdot k_i \gbd_{{#1},i} \right)}
\def\sumk#1{\left( \sum_{i} k _{#1} \cdot k_i \gbd_{{#1},i} \right)}

\def\tn#1#2{t^{[#1]}_{#2}}
\def\L{\left(}\def\R{\right)}
\def\dlips{d{\rm LIPS}}
\def\Split{\mathop{\rm Split}\nolimits}

\def\BR#1#2{\la#1|{K_{abc}}|#2\ra}
\def\BRTTT#1#2{\la#1^+|\Slash{K}_{abc}|#2^+\ra}
\def\BRR#1#2{\la#1|\Slash{K}|#2\ra}
\def\tree{{\rm tree}}
\def\treemhv{{\rm tree\ MHV}}
\def\loopmhv{{\rm 1-loop\ MHV}}
\def\Gr{{\rm Gr}}
\def\nn{\nonumber}
\def\NeqEight{{\cal N} = 8}
\def\NeqFour{{\cal N} = 4}
\def\NeqOne{{\cal N} = 1}
\def\NeqZero{{\cal N} = 0}
\def\ap{\alpha'}
\def\ap{\alpha'}
\def\s(#1,#2){s_{#1,#2}}
\def\F#1#2(#3;#4;#5){ {}_{#1}{\rm F}_{#2}(#3;#4;#5)}
\def\frac#1#2{{#1\over#2}}
\def\tag#1{\eqno{(#1)}}
\def\demo#1{{\sl #1}}
\def\text#1{{ #1}}
\def\ssel{{\rm sl}}
\def\oA#1{{\cal O}({\alpha'}^{#1})}

\newcommand{\ttbs}{\char'134}
\newcommand\fverb{\setbox\fverbbox=\hbox\bgroup\verb}
\newcommand\fverbdo{\egroup\medskip\noindent%
			\fbox{\unhbox\fverbbox}\ }
\newcommand\fverbit{\egroup\item[\fbox{\unhbox\fverbbox}]}
\newbox\fverbbox
\newcommand{\jhepname}{JHEP}

\title{
{Monodromy   and   Kawai-Lewellen-Tye   Relations  for
Gravity
  Amplitudes}}%

\author{
{N. E. J. Bjerrum-Bohr}\\
Niels Bohr International Academy and DISCOVERY Center,\\ The
Niels Bohr Institute, Blegdamsvej 17, \\ DK-2100 Copenhagen \O,
Denmark,\\ {\it bjbohr@nbi.dk}\\ \\ \\
{Pierre Vanhove\thanks{IPHT-T10/023, IHES/P/10/07.}}
\\
Institut des Hautes Etudes Scientifiques, Le Bois-Marie,\\
35 Route de Chartres, F-91440 Bures-sur-Yvette, France\\ and\\
CEA, DSM, Institut de Physique Th{\'e}orique, IPhT, CNRS, MPPU,\\
URA2306, Saclay, F-91191 Gif-sur-Yvette, France,\\
{\it pierre.vanhove@cea.fr}\\ \\ \\
Submitted for the 12th Marcel Grossman Meeting 2009}

\maketitle

\begin{abstract}
We are still learning intriguing new facets of the
 string theory motivated Kawai-Lewellen-Tye (KLT) relations linking
products of amplitudes in Yang-Mills theories and amplitudes in
gravity. This is very clearly displayed in computations of
$\cN=8$ supergravity where the perturbative expansion show a
vast number of similarities to that of $\cN=4$
super-Yang-Mills. We will here investigate how identities based
on monodromy relations for Yang-Mills amplitudes can be very
useful for organizing and further streamlining the KLT
relations yielding even more compact results for gravity
amplitudes.
\\
\\
\\
\\
\\
\\
\\
\\
\\
\\
\\
\\

\end{abstract}



\section{Introduction}
The search for a valid construction of quantum gravity has been
on for most of the previous century initiated by Einstein's
formulation of General Relativity in 1916 and the quantum
mechanics revolution in the 1920ties. Physicists today are
still hunting the answers to the ultimate questions, {\it e.g.}
how was the universe formed and how does one comprehend the
fabric of space and time? Quantum mechanical corrections to
gravity are crucial for the exact answers but the fundamental
concepts of such a quantum theory are unfortunately still very
dim. In this paper we will investigate how we can learn about
an ultimate theory of quantum gravity through studying
symmetries in Yang-Mills theories and the links posed between
Yang-Mills theories and gravity through string theory.

The combination of a traditional quantization and the extra
symmetry introduced by a super-symmetrization of fundamental
interactions appeared for a long while to be a way out of the
troublesome ultraviolet divergences associated with a field
theory for gravity. The most famous model is possibly is the
one of maximal supersymmetry ${\cN=8}$
supergravity~\cite{Cremmer:1978km,ECBJ}. This theory arises as
a  low-energy  effective  description  of  string  theory in
four dimensions. Various later arguments based on supersymmetry
point to a delay in the onset of ultraviolet divergences due to
the extra
symmetry~\cite{Howe:1980th,Green:2006gt,Green:2006yu,Berkovits:2009aw}
but it has long been the belief that only string theory should
be completely free of UV divergences. However since no explicit
ultraviolet divergences have been found so far in the
four-dimensional four-graviton
amplitude~\cite{Green:1982sw,Bern:1998ug,Bern:2007hh,Bern:2007dw,Bern:2006kd,Bern:2008pv},
the effective field-theory status of  ${\cN=8}$ supergravity
and its relation to string theory have been put into questions.

We know that the on-shell $S$-matrix elements in string theory
depend on the scalars parameterizing the (classical) moduli
space $E_{7(7)}(\IR)/(SU(8)/\ZZ_2)$ and that these are
covariant under the discrete U-duality  subgroup
$E_{7(7)}(\ZZ)$~\cite{Hull:1994ys,Green:2010wi}. However in
supergravity the $S$-matrix elements  are invariant under the
continuous symmetry $E_{7(7)}(\IR)$~\cite{ArkaniHamed:2008gz,
Kallosh:2008rr,He:2008pb,Broedel:2009hu}. From the string
theory viewpoint the relation between the four-dimensional
Planck  length $\ell_4$ and the string scale
$\ell_s=\sqrt{\alpha'}$  depend on the (four-dimensional)
dilaton $\ell_4^2=\alpha'\,y_4$ where $y_4=g_s^2\,
{\alpha'}^3/(R_1\cdots R_6)$ and $R_i$ are the radii of
compactification. The decoupling limit of string amplitudes
goes as $\ell_s\to0$, $1/R_i\to\infty$ and
$R_i/\alpha'\to\infty$, keeping the four-dimensional Newton's
constant $2\kappa_{(4)}^2=2\pi\, \ell_4^2$ fixed. This limit is
singular since in this limit some non-perturbative states
become massless and dominate the
$S$-matrix~\cite{Green:2007zzb,Green:2010sp}.  These
non-decoupling results do not imply that $\cN=8$ supergravity
has perturbative ultraviolet problems however because of the
lack of concrete data it has become urgent to clarify the
status of the ultraviolet behavior of $\cN=8$ supergravity in
four dimensions and its relation to string theory.

In recent years, by a combination of different inputs from
string theory, supersymmetry, unitarity and due to remarkable
progress in  computational capacity, a huge number of
amplitudes have been computed~\cite{BernKJ}. Surprisingly the
ultraviolet behavior of $\cN=8$ supergravity occurs explicitly
to be identical to the one of $\cN=4$ super-Yang-Mills at least
through four
loops~\cite{Bern:1998ug,Bern:2005bb,BjerrumBohr:2005xx,
BjerrumBohr:2006yw,BjerrumBohr:2008vc,Bern:2007hh,Bern:2008pv,Bern:2009kf,Green:2006gt,Green:2006yu}.
These results have made it clear that $\cN=8$ supergravity has
a much better perturbative expansion than power-counting
na{\"\i}vely suggests. It is still an open question if the
perturbative expansions of the two theories are similar to all
loop orders or what in given case will be the first loop order
to have a dissimilarity. These and other aspects are discussed
further in ref.~\cite{Vanhove:MG12}.

Motivated by string theory~\cite{Kawai:1985xq} where the
massless spectrum  of $\cN=8$ supergravity can be factorized as
the tensorial product of two copies of $\cN=4$ super-Yang-Mills
theories, one can organize  $\cN=8$ supergravity tree-level
amplitudes according to the KLT
relations~\cite{Kawai:1985xq,Bern:1998ug,BernJI,BernKJ,EffKLT,Bern:2008qj,BjerrumBohr:2010rt}
which we will write schematically in the following way
\begin{equation}\label{e:defKLT}
{\cal M}_{\rm Gravity}^{\rm tree} \sim \sum_{ij} K_{ij} {{\cal A}^{i\; L}_{\rm Yang-Mills}}
\times { {{\cal A}^{j\; R}_{\rm Yang-Mills}}}\,.
\end{equation}
Here $\mathcal{M}_{\rm Gravity}$, ${\cal A}^{i\; L}_{\rm
Yang-Mills}$, ${\cal A}^{j\;R}_{\rm Yang-Mills}$ are gravity
and color ordered Yang-Mills amplitudes and $K_{ij}$ is a
specific function of kinematic invariants needed to ensure that
the tree-level gravity amplitude has the  correct analytic
structure.

The simple KLT relations between theories of gravity and two
gauge theories are observed directly in on-shell $S$-matrix
elements but have no motivation at the Lagrangian level (This
is true even if part of the Lagrangian is rearranged as a
product of Yang-Mills types of interactions at the
two-derivative
level~\cite{Ananth:2007zy,Ananth:2009kn,BjerrumBohr:2010rt} or
for higher derivative corrections~\cite{Peeters:2000qj}. In the
case of pure gravity one needs to take into account the
contribution from the dilaton in employing the KLT relations.)

Because of their high degree of supersymmetry both $\cN=4$
super-Yang-Mills and $\cN=8$ supergravity loop amplitudes are
cut constructible in $D=4-2\epsilon$ dimensions and
surprisingly the knowledge of the tree-level amplitudes is
enough for reconstructing the full higher-loop
amplitudes~\cite{Bern:2007xj,Bern:2007hh,Bern:2007dw,ArkaniHamed:2008gz,Bern:2008pv}.

We will here discuss tree amplitudes from the point of view of
the classical $\cN=8$ theory, which can be constructed from the
$\cN=4$ super-Yang-Mills tree-level amplitudes using the KLT
relation in~(\ref{e:defKLT}). (For effective theories of
gravity~\cite{EffKLT} one can also employ KLT relations in a
slightly modified fashion taking into account   higher
derivative   operators  introduced through counterterms to
ultraviolet divergences.)

We  will  next discuss  the  construction  of tree-level
amplitudes  in Yang-Mills and gravity from a minimal basis of
amplitudes following~\cite{BjerrumBohr:2009rd}.

\section{Minimal basis for Yang-Mills and Gravity tree-level amplitudes}
The $n$-point amplitude in open string theory with $U(N)$ gauge
group reads
\begin{eqnarray}
\label{e:AmpNdef}
\cA_n =ig_{\rm YM}^{n-2}\, (2\pi)^{D}\,\delta^{D}(k_1+\cdots+k_n)\hspace{-0.7cm}
\sum_{(a_1,\dots,a_n)\in S_n/\ZZ_n}\hspace{-0.7cm} \tr(T^{a_1}
\cdots T^{a_n})\, \cA_n(a_1,\cdots,a_n)\,,
\end{eqnarray}
where  $D$ is any number of dimensions obtained by dimensional
reduction from $26$ dimensions if we consider the bosonic
string, or $10$ dimensions in the supersymmetric case. The
field theory amplitudes are obtained by taking the limit
$\alpha'\to0$. A new series of  amplitude identities between
different color-ordered amplitudes based on monodromy for
integrations in string theory was derived
in~\cite{BjerrumBohr:2009rd}
(see~\cite{Stieberger:2009hq,Mafra:2009bz} for  related
discussions). The real part of these  relations relates the
$n$-point amplitude with different orderings as
\begin{align}\label{stringKK}
&{\cal A}_n(\beta_1,\ldots,\beta_r,1,\alpha_1,\ldots,\alpha_s,n)
=(-1)^r \nonumber \\
&\phantom{AAAAAAAAA}\times \Re{\rm e}\Big[\!\!\!\!\!\!
\prod_{1\leq i <j \leq r}\!\!\!\!\!\!\!e^{2i\pi\alpha'(k_{\beta_i}\cdot
k_{\beta_j})}\!\!\!\!\!\!\!\!\!\!\!\!\!\!
\sum_{\sigma\subset{\rm OP}
\{\alpha\}\cup\{\beta^T\}}\!\prod_{i=0}^s
\prod_{j=1}^r\! e^{(\alpha_i,\beta_j)}{\cal A}_n(1,\sigma,n)\Big]\,.\
\end{align}
Here $e^{({\alpha,\beta})} \equiv e^{2i\pi\alpha'(k_\alpha\cdot
k_\beta)}$ if  $x_{\beta} >  x_{\alpha}$  and 1  otherwise,
$\alpha_0$ denotes the leg 1 at point 0.
The imaginary part give the following amplitude relation
\begin{equation}\begin{split}\label{stringBern}
0=\hspace{-0cm}\Im{\rm m}\Big[\!\!\!\!\!\!
\prod_{1\leq i <j \leq r}\!\!\!\!\!\!
\!e^{2i\pi\alpha'(k_{\beta_i}\cdot k_{\beta_j})}\!\!\!\!\!\!\!\!\!\!\!\!\!\!
\sum_{\sigma\subset{\rm OP}
\{\alpha\}\cup\{\beta^T\}}\!\prod_{i=0}^s\prod_{j=1}^r
\!e^{(\alpha_i,\beta_j)}\cA_n(1,\sigma,n)\Big]\,.\vspace{-0.3cm}\
\end{split}\end{equation}\noindent
We define the $(n-3)!$ color ordered amplitudes $\cB_\sigma=
\cA_n(1,\sigma(2),\cdots,\sigma(n-2),n-1,n)$ with $\sigma \in
\mathfrak S_{n-3}$ denoting a permutation of the legs
$(2,\dots,n-2)$. As a consequence of~(\ref{stringKK})
and~(\ref{stringBern}) any color ordered amplitudes associated
with the permutation $\sigma'$ of the external legs can be
expanded~\cite{BjerrumBohr:2009rd}\vspace{-0.1cm}
\begin{equation}
\label{eq:1}
\cA_n(\sigma'(1),\cdots,\sigma'(n))=   \sum_{\sigma\in\mathfrak  S_n}  \,
c^\sigma_{\sigma'} \, \cB_\sigma\,,
\vspace{-0.1cm}\end{equation}
where $c^\sigma_{\sigma'}$ are functions of the
$\cSS{p}{q}=\sin(2\pi\alpha'\,p\cdot q)$ and $p$ and $q$ are
sums of the external momenta. This implies that
$\{\cB_\sigma;\sigma\in\mathfrak S_{n-3}\}$ provides a minimal
basis in which all other color ordered amplitudes can be
expanded.

Because the monodromy relations hold for all polarization
configurations and any smaller number of dimensions by a
trivial dimensional reduction, it follows immediately that they
hold for any choice of external legs corresponding to the full
${\cal N} = 1$,  $D=10$ supermultiplet and in dimensional
reductions thereof~\cite{Sondergaard:2009za}.

In the case of the four-gluon amplitude one have
\begin{eqnarray}
\cA_4(1,2,3,4)&=&{\Gamma(1-\alpha's)\Gamma(1-\alpha't)\over\Gamma(1-\alpha'u)}\,
\left( {n_s\over s}+{n_t\over t}\right)\,,\\
\cA_4(1,3,2,4)&=&{\Gamma(1-\alpha'u)\Gamma(1-\alpha't)\over\Gamma(1-\alpha's)}\,
\left( -{n_u\over u}-{n_t\over t}\right)\,,\\
\cA_4(2,1,3,4)&=&{\Gamma(1-\alpha's)\Gamma(1-\alpha'u)\over\Gamma(1-\alpha't)}\,
\left( {n_s\over s}+{n_u\over u}\right)\,,
\end{eqnarray}
where  $n_s$, $n_t$  and $n_u$  depends on  the polarizations
and the external momenta.

The monodromy relations~(\ref{stringKK}) and~(\ref{stringBern})
\begin{equation}
\cA_4(1,3,2,4)\!=\!{\sin(2\pi\alpha'\,s)\over\sin(2\pi\alpha'\,u)}\cA_4(1,2,3,4)\,,\
\
\cA_4(2,1,3,4)\!=\!{\sin(2\pi\alpha'\,t)\over\sin(2\pi\alpha'\,u)}\cA_4(1,2,3,4)\,,
\end{equation}
imply that the numerator factors satisfy the Jacobi like
relation $n_s=n_t+n_u$. The generalization to higher points
gives the new amplitude relations recently conjectured by Bern
et al. in ref.~\cite{Bern:2008qj}.  The string theory monodromy
identities for the Kawai-Lewellen-Tye relationship between
closed and open string amplitudes give highly symmetric forms
for tree-level amplitudes where the tree-level gravity
amplitudes are expanded in a basis obtained by the left/right
tensorial product of gauge color ordered amplitudes
\begin{equation}
\cM_n=\sum_{\sigma,\sigma'\in\mathfrak S_{n-3}}\,
\cG^{\sigma,\sigma'}(k_i\cdot k_j)\, \cB_\sigma^L \cB_{\sigma'}^R\, .
\end{equation}
As a direct application of our procedure, we can rewrite the
Kawai-Lewellen-Tye relations at four-point level as
\begin{equation}
\cM_4={\kappa_{(4)}^2\over
\alpha'}\,{\cSS{k_1}{k_2}\cSS{k_1}{k_4}\over\cSS{k_1}{k_3}}\,\cA_4^L(1,2,3,4)
\cA_4^R(1,2,3,4)\,.
\label{KLTFour}
\end{equation}
The field theory limit of  the  string
amplitude~(\ref{KLTFour}), $\alpha'\to0$  gives  the symmetric
form of the gravity amplitudes of~\cite{Bern:2008qj}
\begin{equation}
\cM_4^{FT}   =\kappa_{(4)}^2\,{st\over   u}\,\left({n_s\over  s}+{n_t\over
t}\right)\left({\tilde n_s\over s}+{\tilde n_t\over
t}\right)
=-\kappa_{(4)}^2\,\left({n_s\tilde n_s\over s}+{n_t\tilde n_t\over
t}+{n_u\tilde n_u\over u}\right) \,.
\label{KLTFourFT}
\end{equation}
Here we have made use of the on-shell relation $s+t+u=0$ and
the four-point Jacobi relation $n_u=n_s-n_t$. Similarly
considerations at higher-point order will be detailed
in~\cite{BDSVtoappear}.

\section{Conclusions}
We have discussed the interesting link posed by the
Kawai-Lewellen-Tye (KLT) string theory relations between
products of amplitudes in Yang-Mills theories and amplitudes in
gravity. We here observed how identities based on monodromy
relations for Yang-Mills amplitudes and the KLT relations can
be employed to yield very compact results for gravity
amplitudes. It would be interesting to analyze the role of the
monodromies at loop order since this would allow us to further
understand the similarities of the perturbative expansion of
$\cN =8$ supergravity and $\cN=4$ super Yang-Mills.

\section{Acknowledgments}
(NEJBB) is Knud H{\o}jgaard Assistant Professor at the Niels
Bohr International Academy.


\begin{thebibliography}{99}
\bibitem{Cremmer:1978km}
  E.~Cremmer, B.~Julia and J.~Scherk,
  Phys.\ Lett.\  B {\bf 76}, 409 (1978);
  E.~Cremmer and B.~Julia,
  Phys.\ Lett.\  B {\bf 80}, 48 (1978).

\bibitem{ECBJ} E.~Cremmer and B.~Julia,
Nucl.\ Phys.\  B {\bf 159}, 141 (1979).

\bibitem{Howe:1980th} P.~S.~Howe and U.~Lindstrom,
Nucl.\ Phys.\  B {\bf 181} (1981) 487;
R.~E.~Kallosh,
Phys.\ Lett.\  B {\bf 99} (1981) 122;
P.~S.~Howe, K.~S.~Stelle and P.~K.~Townsend,
Nucl.\ Phys.\  B {\bf 236} (1984) 125;
P.~S.~Howe and K.~S.~Stelle,
Phys.\ Lett.\  B {\bf 554} (2003) 190 [hep-th/0211279].

\bibitem{Green:2006gt} M.~B.~Green, J.~G.~Russo and P.~Vanhove,
JHEP {\bf 0702} (2007) 099.

\bibitem{Green:2006yu} M.~B.~Green, J.~G.~Russo and P.~Vanhove,
Phys.\ Rev.\ Lett.\  {\bf 98} (2007) 131602 [hep-th/0611273].

\bibitem{Berkovits:2009aw}
N.~Berkovits, M.~B.~Green, J.~G.~Russo and P.~Vanhove,
JHEP {\bf 0911} (2009) 063 [0908.1923 [hep-th]].

\bibitem{Green:1982sw}
M.~B.~Green, J.~H.~Schwarz and L.~Brink,
Nucl.\ Phys.\  B {\bf 198} (1982) 474.

\bibitem{Bern:1998ug} Z.~Bern, L.~J.~Dixon, D.~C.~Dunbar,
    M.~Perelstein and J.~S.~Rozowsky,
Nucl.\ Phys.\  B {\bf 530}, 401 (1998) [hep-th/9802162].

\bibitem{Bern:2007hh} Z.~Bern,  J.~J.~Carrasco,  L.~J.~Dixon,
    H.~Johansson,  D.~A.~Kosower and R.~Roiban,
Phys.\ Rev.\ Lett.\  {\bf 98} (2007) 161303 [hep-th/0702112].

\bibitem{Bern:2007dw}
Z.~Bern, L.~J.~Dixon and D.~A.~Kosower,
Annals Phys.\  {\bf 322} (2007) 1587 [0704.2798 [hep-ph]].

\bibitem{Bern:2006kd} Z.~Bern, L.~J.~Dixon and R.~Roiban,
Phys.\ Lett.\  B {\bf 644} (2007) 265 [hep-th/0611086].

\bibitem{Bern:2008pv} Z.~Bern, J.~J.~M.~Carrasco, L.~J.~Dixon,
    H.~Johansson and R.~Roiban,
Phys.\ Rev.\  D {\bf 78} (2008) 105019 [0808.4112 [hep-th]].


\bibitem{Hull:1994ys}
C.~M.~Hull and P.~K.~Townsend,
Nucl.\ Phys.\  B {\bf 438} (1995) 109
[arXiv:hep-th/9410167].



\bibitem{Green:2010wi}
M.~B.~Green, J.~G.~Russo and P.~Vanhove,
1001.2535 [hep-th].

\bibitem{ArkaniHamed:2008gz}
N.~Arkani-Hamed, F.~Cachazo and J.~Kaplan,
0808.1446 [hep-th].

\bibitem{Kallosh:2008rr}
R.~Kallosh and T.~Kugo,
JHEP {\bf 0901} (2009) 072 [0811.3414 [hep-th]].

\bibitem{He:2008pb}
S.~He and H.~Zhu,
0812.4533 [hep-th].

\bibitem{Broedel:2009hu}
J.~Broedel and L.~J.~Dixon,
0911.5704 [hep-th].

\bibitem{Green:2007zzb} M.~B.~Green, H.~Ooguri and
    J.~H.~Schwarz,
Phys.\ Rev.\ Lett.\  {\bf 99}, 041601 (2007).

\bibitem{Green:2010sp}
M.~B.~Green, J.~G.~Russo and P.~Vanhove,
1002.3805 [hep-th].

\bibitem{BernKJ} Z.~Bern,
Living Rev.\ Rel.\  {\bf 5}, 5 (2002) [gr-qc/0206071].

\bibitem{Bern:2005bb} Z.~Bern, N.~E.~J.~Bjerrum-Bohr and
    D.~C.~Dunbar,
JHEP {\bf 0505}, 056 (2005) [hep-th/0501137].

\bibitem{BjerrumBohr:2005xx} N.~E.~J.~Bjerrum-Bohr,
    D.~C.~Dunbar and H.~Ita,
Phys.\ Lett.\  B {\bf 621}, 183 (2005) [hep-th/0503102].

\bibitem{BjerrumBohr:2006yw} N.~E.~J.~Bjerrum-Bohr,
    D.~C.~Dunbar, H.~Ita, W.~B.~Perkins and K.~Risager,
JHEP {\bf 0612} (2006) 072 [hep-th/0610043].

\bibitem{BjerrumBohr:2008vc} N.~E.~J.~Bjerrum-Bohr and
    P.~Vanhove,
JHEP {\bf 0804} (2008) 065 [0802.0868 [hep-th]].

\bibitem{Bern:2009kf} Z.~Bern, J.~J.~M.~Carrasco and
    H.~Johansson,
0902.3765 [hep-th].

\bibitem{Vanhove:MG12} P.~Vanhove,  
IHES-P/10/02, IPHT-T-09-190

\bibitem{Kawai:1985xq} H.~Kawai, D.~C.~Lewellen and
    S.~H.~H.~Tye,
Nucl.\ Phys.\  B {\bf 269} (1986) 1.

\bibitem{BernJI} Z.~Bern and A.~K.~Grant,
Phys.\ Lett.\  B {\bf 457}, 23 (1999) [hep-th/9904026];
S.~Ananth and S.~Theisen,
Phys.\ Lett.\  B {\bf 652}, 128 (2007) [0706.1778 [hep-th]].
N.~E.~J.~Bjerrum-Bohr and O.~T.~Engelund,
1002.2279 [hep-th].

\bibitem{EffKLT} Z.~Bern, A.~De Freitas and H.~L.~Wong,
Phys.\ Rev.\ Lett.\  {\bf 84}, 3531 (2000) [hep-th/9912033];
N.~E.~J.~Bjerrum-Bohr,
Phys.\ Lett.\ B {\bf 560}, 98 (2003) [hep-th/0302131];
Nucl.\ Phys.\ B {\bf 673}, 41 (2003) [hep-th/0305062];
N.~E.~J.~Bjerrum-Bohr and K.~Risager,
Phys.\ Rev.\ D {\bf 70}, 086011 (2004) [hep-th/0407085].

\bibitem{Bern:2008qj} Z.~Bern, J.~J.~M.~Carrasco and
    H.~Johansson,
Phys.\ Rev.\  D {\bf 78}, 085011 (2008) [0805.3993 [hep-ph]].

\bibitem{BjerrumBohr:2010rt}
N.~E.~J.~Bjerrum-Bohr and O.~T.~Engelund,
1002.2279 [hep-th].

\bibitem{Ananth:2007zy} S.~Ananth and S.~Theisen,
Phys.\ Lett.\  B {\bf 652} (2007) 128 [0706.1778 [hep-th]].

\bibitem{Ananth:2009kn} S.~Ananth,
0902.3128 [hep-th].

\bibitem{Peeters:2000qj} K.~Peeters, P.~Vanhove and
    A.~Westerberg,
Class.\ Quant.\ Grav.\  {\bf 18} (2001) 843 [hep-th/0010167].

\bibitem{Bern:2007xj}
Z.~Bern, J.~J.~Carrasco, D.~Forde, H.~Ita and H.~Johansson,
Phys.\ Rev.\  D {\bf 77} (2008) 025010 [0707.1035 [hep-th]].


\bibitem{BjerrumBohr:2009rd}
N.~E.~J.~Bjerrum-Bohr, P.~H.~Damgaard and P.~Vanhove,
Phys.\ Rev.\ Lett.\  {\bf 103} (2009) 161602 [0907.1425
[hep-th]].

\bibitem{Stieberger:2009hq}
S.~Stieberger,
0907.2211 [hep-th].

\bibitem{Mafra:2009bz}
C.~R.~Mafra,
JHEP {\bf 1001} (2010) 007 [0909.5206 [hep-th]].

\bibitem{Sondergaard:2009za} T.~S{\o}ndergaard,
Nucl.\ Phys.\  B {\bf 821} (2009) 417 [0903.5453 [hep-th]].

\bibitem{BDSVtoappear}  N.~E.~J.~Bjerrum-Bohr,  P.~H.Damgaard,
    T.~S{\o}ndergaard and P.~Vanhove, hep-th/0003530.

\end{thebibliography}
\end{document}